\documentclass[twocolumn,superscriptaddress,amssymb,amsmath,nobibnotes,aps,
prd,showpacs,nofootinbib]{revtex4}%
\pdfoutput=1

\usepackage{graphicx}
\usepackage{epsf}
\usepackage{bm}
\usepackage{amsmath}
\usepackage{amsfonts}
\usepackage{amssymb}
\usepackage{epstopdf}
\usepackage{natbib}%
\usepackage{rotating}
\usepackage{pdflscape}
\usepackage{adjustbox}
\usepackage[toc,page]{appendix}
\providecommand{\U}[1]{\protect\rule{.1in}{.1in}}
\newcommand{\be}{\begin{equation}}
\newcommand{\ee}{\end{equation}}

\newcommand{\mincir}{\raise
-3.truept\hbox{\rlap{\hbox{$\sim$}}\raise4.truept\hbox{$<$}\ }}
\newcommand{\magcir}{\raise
-3.truept\hbox{\rlap{\hbox{$\sim$}}\raise4.truept\hbox{$>$}\ }}

\usepackage{color}

\begin{document}

\title{New constraints on interacting dark energy from cosmic chronometers}

\author{Rafael C. Nunes}
\email{nunes@ecm.ub.edu}
\affiliation{Departamento de F\'isica, Universidade Federal de Juiz de Fora, 36036-330,
Juiz de
Fora, MG, Brazil}

\author{Supriya Pan}
\email{span@iiserkol.ac.in}
\affiliation{Department of Physical Sciences, Indian Institute of Science Education and
Research -- Kolkata, Mohanpur -- 741246, West Bengal, India}

\author{Emmanuel N. Saridakis}
\email{Emmanuel\_Saridakis@baylor.edu}
\affiliation{Instituto de F\'{\i}sica, Pontificia Universidad de Cat\'olica de
Valpara\'{\i}so,
Casilla 4950, Valpara\'{\i}so, Chile}
\affiliation{CASPER, Physics Department, Baylor University, Waco, TX 76798-7310, USA}

\pacs{98.80.-k, 95.36.+x, 95.35.+d, 98.80.Es}

\begin{abstract}
We use the latest compilation of observational Hubble parameter measurements estimated
with the differential evolution of cosmic chronometers, in combination with the local
value of the Hubble constant recently measured with 2.4$\%$ precision, to constrain
the cosmological scenario where dark energy interacts directly with the dark matter
sector. To diminish the degeneracy between the parameters we additionally consider
standard probes, such as Supernovae Type Ia from joint light curves (JLA)
sample, Baryon Acoustic Oscillation distance measurements (BAO), and cosmic 
microwave background data from Planck 2015 estimations. 
Our analysis shows that the direct interaction between dark energy and dark matter is 
mildly flavored, while the dark energy equation-of-state parameter is $w < - 
1$ at 3$\sigma$ confidence level. 
\end{abstract}

\maketitle

\section{Introduction}
\label{sec:intro}

According to latest observations \cite{Planck2015}, around 69\% of the universe is
constituted by the dark energy (DE) sector, responsible for its current
accelerating phase, and almost 26\% by the cold dark matter (CDM) sector in the form of
dust. These two sectors are considered not to interact with each other, and the
resulting scenario has been believed to be the best description of the present observable
universe, due to its agreement with a large number of independent observations. However,
some potential problems such as the coincidence one \cite{Steinhardt2003}, namely why are
the dark energy and dark matter energy densities of the same order although they follow
completely different evolutions, have led to a large amount of research towards scenarios
in which the above two dark sectors exhibit a direct interaction. Such a mutual direct
interaction between dark energy and dark matter cannot be excluded from the field
theoretical point of view, independently of the specific nature of the former, i.e. 
whether
it arises through a universe-content modification \cite{Copeland:2006wr,Cai:2009zp} or
through a gravitational modification \cite{Nojiri:2010wj,Capozziello:2011et}. Hence, many
different scenarios of interacting dark energy have been constructed (see
\cite{DE_DM_1,DE_DM_2} for reviews and references therein). Interestingly enough,
recently it has been shown that the current data might favor the late-time interaction
between CDM and the vacuum energy 
\cite{Salvatelli:2014zta,Sola:2016_b,Sola:2016jky,Sola:2015} or 
the 
dynamical
dark energy sector
\cite{He:2010im,Xu:2013jma,Costa:2013sva,Ruiz:2014hma,Cai:2015zoa,Li:2015vla,
Richarte:2015maa,Valiviita:2015dfa, Eingorn:2015rma,Pan:2012ki,Murgia:2016ccp}.

In the present work we are interested in providing updated constraints on the scenarios
of interacting dark energy, using the very recent observational data, along with the new
local value of the  Hubble parameter with a 2.4$\%$ determination, released in
\cite{riess}. In particular, considering a direct interaction between dark energy and
dark matter in a Friedmann-Robertson-Walker (FRW) background, and assuming that both
(effective) fluids obey a barotropic equation of state,  the cosmological dynamics becomes
richer and thus the evolution equations for CDM and dark energy become different from
their standard forms. Hence, one can use observational data in order to fit the deviation
from non-interaction, without the need to specify a specific interaction form, i.e.
keeping the analysis in a general ground. The manuscript is organized  as follows: In
Section \ref{sec:ide} we provide a brief description of the expansion history of the
universe in the background of an interacting scenario. In Section \ref{data} we describe
the latest data sets for our analysis, while in Section \ref{results} we use them
and we extract the constraints on the various quantities. Finally, Section \ref{discu}
is devoted to summary and conclusions.

\section{Interacting dark energy}
\label{sec:ide}

Let us briefly describe the scenario of interacting dark energy. In a background of a
FRW universe with metric $ds^2=-dt^2+a(t)^2[dr^2/(1-kr^2)+r^2(d\theta^2+\sin^2\theta
\,d\phi^2)]$, where $a(t)$ is the scale factor and $k$ the
spatial curvature (with $k=0,-1,+1$ for flat, open and closed universe
respectively), the first
Friedmann equation can be written as
\begin{align}
\label{friedmann}
H^2+ \frac{k}{a^2} & = \frac{8 \pi G}{3} \left(\rho_{\gamma} + \rho_b + \rho_{cdm} +
\rho_{de}  \right),
\end{align}
where $H= \dot{a}/a$ is the Hubble rate, with dots denoting
derivatives with respect to the cosmic time, and with $\rho_{\gamma}$, $\rho_b$,
$\rho_{cdm}$, and $\rho_{de}$ denoting the energy densities of radiation, baryons, cold
dark matter and dark energy, respectively.

Since the physics of the radiation and baryonic sector is known quite well, we assume
that these sectors are independently conserved. On the other hand, if the dark energy and
dark matter sectors interact directly, the
energy conservation law for the interacting CDM-DE components reads as $u_{\alpha}
T^{\alpha
\beta}_{;
\beta} = 0$, where the total energy-momentum tensor is
$T^{\alpha \beta}= T^{\alpha \beta}_{cdm} + T^{\alpha \beta}_{de}$, or equivalently
\cite{DE_DM_1,DE_DM_2}
\begin{align}
\dot{\rho}_{cdm} + 3\frac{\dot{a}}{a}\rho_{cdm} = -\dot{\rho}_{de} -
3\frac{\dot{a}}{a}\left(\rho_{de} +
p_{de}\right)=Q,
\label{dotrho}
\end{align}
where $Q$ is a specific interaction form that has to be chosen by hand.
As usual, one can quantify the effect of the above interaction on the dark matter
evolution through a deviation from the standard dust-matter evolution, namely considering
that the dark matter energy density evolves as
\cite{coupled01,coupled02,coupled03,Chen:2011cy}
\begin{align}
\rho_{cdm} = \rho_{cdm,0}\,a^{-3 + \delta},
\label{rho0a3}
\end{align}
with $\rho_{cdm,0}$ the present value, and where the parameter $\delta$ becomes zero in
the non-interacting case. Hence, the parameter $\delta$ quantifies the
deviation from non-interacting case ($\delta<0$ corresponds to energy flow from
dark matter to dark energy), without the need to consider a specific interaction form
$Q$. Combining (\ref{dotrho}), (\ref{rho0a3}) and considering that the dark energy sector
is described by an equation-of-state parameter of the form $w =  p_{de}/\rho_{de}$, we
find that the energy density of the dark energy component evolves as
\begin{align}
\rho_{de} = \rho_{de,0}\,a^{-3(1+w)} + \frac{\delta\, \rho_{cdm,0}}{3|w| - \delta}\left[ 
a^{-3 +\delta}- a^{-3(1+w)}
\right],
\end{align}
where $\rho_{de,0}$ is the present value 
of $\rho_{de}$. Clearly, in the absence of a coupling with the CDM component, the 
conventional dynamical dark energy scenario is recovered. Note that for $w = -1$ 
and $\delta \neq 0$,
we may identify $\rho_{de} = \rho_{vacuum}$
and the above expressions reduce to the vacuum decaying scenario
\cite{coupled1,coupled1b,coupled2,coupled3,coupled4,coupled5,coupled6}.

\section{Current Observational Constraints}
\label{data}

In  the present  work  we   confront  the  constraints  on the cosmological 
parameters
that  can  be  obtained  by probes that map the late-time universe ($z < 2.36$) expansion
history. The baseline of our analysis is the Hubble parameter measurements obtained with
the cosmic chronometers (CC) technique, however we additionally consider standard
probes such as Supernovae Type Ia (SNe Ia), local Hubble parameter value $H_0$ ones, and
Baryon Acoustic Oscillation distance measurements (BAO), to diminish the
degeneracy between the free parameters of the models. In the following subsections, we
present the data sets considered in our analysis.

\subsection{Cosmic chronometer dataset and local value of the Hubble constant}
\label{cc-data}

The CC approach to measure $H(z)$ was first introduced in \cite{cc1}, and it uses relative
ages of the most massive and passively evolving galaxies to measure $dz/dt$, from which
$H(z)$ is inferred. The  latest  implementation  has  been  explained  in  detail  in
\cite{cc2}, where  the possible  sources  of uncertainty  and  related  issues  are  also
discussed. We consider the compilation of Hubble parameter measurements provided by
\cite{cc3, cc2}. It contains the latest updated list of $H(z)$ measurements
\cite{cc2,cc3,cc4,cc5,cc6} obtained with the cosmic chronometers approach,
comprising of 30 measurements spanning the redshift range $0 < z < 2$. This sample covers
roughly 10 Gyr of cosmic time. Furthermore, in our analysis we also include the new local
value of $H_0$ as measured by \cite{riess} with a 2.4 $\%$ determination, which yields
$H_0 = 73.02 \pm 1.79$ km/s/Mpc.

\subsection{Type Ia Supernovae}
\label{snia-data}

SNe Ia were the main tool to discover late-time acceleration \cite{snia1,snia2}, and
they still provide the best constraints on dark energy sector. SNe Ia are very bright
``standard candles'' and thus are used to measure cosmic distances. We consider here the
latest ``joint light curves" (JLA) sample \cite{snia3}, comprised of 740 SNe Ia
in the redshift range $z \in [0.01, 1.30]$. From the observational point of view, the
distance modulus of a SNe Ia can be abstracted from its light curve assuming that
supernovae with identical color, shape and galactic environment have on average the same
intrinsic luminosity for all redshifts. This hypothesis is quantified by a
empirical linear relation, yielding a standardized distance modulus $\mu = 5
\log_{10}(d_L/10pc)$ of the form
\begin{align}
\label{snia1}
 \mu = m^\ast_B - (M_B - \alpha \times X_1 + \beta \times C),
\end{align}
where $m^\ast_B$ corresponds to the observed peak magnitude in rest frame B band and
$\alpha$, $\beta$, and $M_B$ are nuisance parameters in the distance estimate.
The absolute magnitude is related to the host stellar mass ($M_{stellar}$) by a simple
step function:
$M_B = M_B$ if $M_{stellar} < 10^{10} M_\odot $, otherwise $M_B = M_B + \Delta_M$. The
light-curve parameters ($m^\ast_B$, $X_1$, and $C$) result from the fit of a model of 
the
SNe Ia spectral sequence to the photometric data. In our analysis we assume $M_B$,
$\Delta_M$, $\alpha$ and $\beta$ as nuisance parameters.

\subsection{Baryon Acoustic oscillation}
\label{bao-data}

Another important cosmological probe are the baryon acoustic oscillations (BAO). The BAO
can be traced to pressure waves at the recombination epoch, generated by cosmological
perturbations in the primeval baryon-photon plasma, appearing as distinct peaks in
large angular scales. We  use  the  following  BAO  data  to  constrain  the  expansion
history of the scenario at hand: the  measurement from  the  Six Degree  Field  Galaxy
Survey (6dF) \cite{bao1}, the  Main Galaxy Sample  of  Data  Release 7  of  Sloan
Digital  Sky  Survey  (SDSS-MGS) \cite{bao2}, the  LOWZ  and  CMASS  galaxy  samples  of
the Baryon  Oscillation  Spectroscopic  Survey  (BOSS-LOWZ  and  BOSS-CMASS,
respectively) \cite{bao3},  and the distribution of the LymanForest in BOSS (BOSS-Ly)
\cite{bao4}. These measurements and their corresponding effective redshift $z$ are
summarized in Table \ref{tab1}.
\begin{table}[!h]
      \begin{center}
          \begin{tabular}{ccccc}
          \hline
          \hline
           Survey &  $z$     &  Parameter   &  Measurement  & Reference  \\
          \hline
 6dF             & 0.106 & $r_s/D_V$  &  0.327 $\pm$ 0.015 & \cite{bao1} \\
 SDSS-MGS        & 0.10  &  $D_V/r_s$ &  4.47 $\pm$ 0.16   & \cite{bao2} \\
 BOSS-LOWZ       & 0.32  & $D_V/r_s$  &  8.47  $\pm$ 0.17  & \cite{bao3} \\
 BOSS-CMASS      & 0.57  &  $D_V/r_s$ &  13.77  $\pm$ 0.13 & \cite{bao3} \\
 BOSS-$Ly_{\alpha}$& 2.36& $c/(H r_s)$&  9.0    $\pm$ 0.3  & \cite{bao4} \\
 BOSS-$Ly_{\alpha}$& 2.36 & $D_A/r_s$ &  10.08  $\pm$ 0.4  & \cite{bao4} \\
          \hline
          \hline
          \end{tabular}
      \end{center}
      \caption{Baryon acoustic oscillation (BAO) data measurements included in our
analysis.}
      \label{tab1}
\end{table}

\subsection{Cosmic microwave background}
\label{cmb-data}

Cosmic microwave background (CMB) radiation provides a unique window to understand the 
structure formation in a given cosmological model. In order to avoid the 
calculation of the evolution of the linear density perturbations, and incorporate as much 
empirical information as possible, one can develop a substitution for the full 
Boltzmann analysis of CMB. The fundamental principle uses certain characteristic 
distance scales to summarize the CMB data, namely the shift parameter $R$ 
\cite{cmb1,cmb2} that determines the amplitude of acoustic peaks in 
the power spectrum of CMB temperature anisotropy, and the acoustic scale $l_a$ that 
determines the acoustic peak structure \cite{cmb3}. In this work we use the distance 
measurements $R$ and $l_a$, obtained in \cite{cmb4} by using the Markov Chain Monte Carlo 
chains from Planck TT, TE, EE $+$ lowP data in Planck Legacy Archive. We 
use the measurements marginalized over the amplitude of the lensing power spectrum.

Let us now describe how a coupled dark energy affects the anisotropies of the cosmic 
microwave background radiation and the matter spectrum. We follow 
\cite{delta_cdm,delta_cdm2}, where a synchronous gauge has been 
adopted, and thus the line element of the linearly perturbed FRW metric can 
be written as
\begin{eqnarray}
ds^2 = -dt^2 + a^2(t)(\delta_{ij} + h_{ij})dx^idx^j,
\end{eqnarray}
where $h_{ij}$ denotes the metric perturbation. For simplicity we restrict to the 
scalar modes $h$ and $\eta$ of the metric perturbations, where $h$, $\eta$ are 
respectively the trace and traceless parts of 
the metric $h_{ij}$, which in 
Fourier space read as
\begin{eqnarray}
&&\!\!\!\!\!\!\!\!\!\!\!\!\!\!\!\!\!\!
h_{ij}(x,\tau) = \int d^3k \exp^{i\vec{k} \cdot \vec{x}} \Big[ \hat{k}_i \hat{k}_j 
h(\vec{k},\tau)
\nonumber \\
&&
\ \ \ \ \ \ \ \ \ \ \ \ \ \ \ \ \ \ \ \ \ \ +
\Big(k_i k_j - \frac{1}{3} \delta_{ij}\Big) 6\eta(\vec{k},\tau) \Big], 
\end{eqnarray}
with $k$  the wavenumber of the Fourier mode ($\vec{k} = k \hat{k}$). 
We assume that the dark energy perturbation in the dark matter comoving frame is 
identically zero. Hence, in an interacting dark energy - dark matter system, the 
perturbed part of the energy momentum for the dark matter evolution writes as
\cite{delta_cdm,delta_cdm2}:
\begin{eqnarray}
\label{delta_dm}
\dot{\delta}_{cdm} - \frac{Q}{\rho_m}\delta_{cdm} - \frac{k^2}{a^2}\theta_{cdm} + 
\frac{\dot{h}}{2} 
= 0,
\end{eqnarray}
where ${\delta}_{cdm}$ is  the total density perturbation
of cold dark matter, with  $\theta_{cdm}$ its  total covariant
velocity perturbation, while the interaction term $Q$   can be obtained through 
(\ref{dotrho}) and  (\ref{rho0a3}).

As usual we work in a comoving synchronous gauge, in which the dark 
matter velocity is zero. The baryon component is conserved 
independently, and the perturbation equations for the baryon density contrast and 
velocity respectively write as 
$\dot{\delta_{b}} - \frac{k^2}{a^2}\theta_{b} = - \frac{\dot{h}}{2}$ and 
$\dot{\theta}_{b} 
= 0$. Finally, photons and neutrinos are also conserved 
and follow the standard evolution equation \cite{Ma_Bertschinger}. 
 
\section{Results}
\label{results}

We now proceed to the main analysis, namely to use the above data sets in order to
constrain the scenario of interacting dark energy described in Section \ref{sec:ide}. In
order to fit the free parameters of the scenario we use the public code CLASS \cite{class}
in interface with the public Monte Carlo code Monte Python \cite{monte}. Moreover, we
choose the Metropolis Hastings algorithm as our sampling method.

\begin{figure}
	\includegraphics[width=3.3in, height=2.7in]{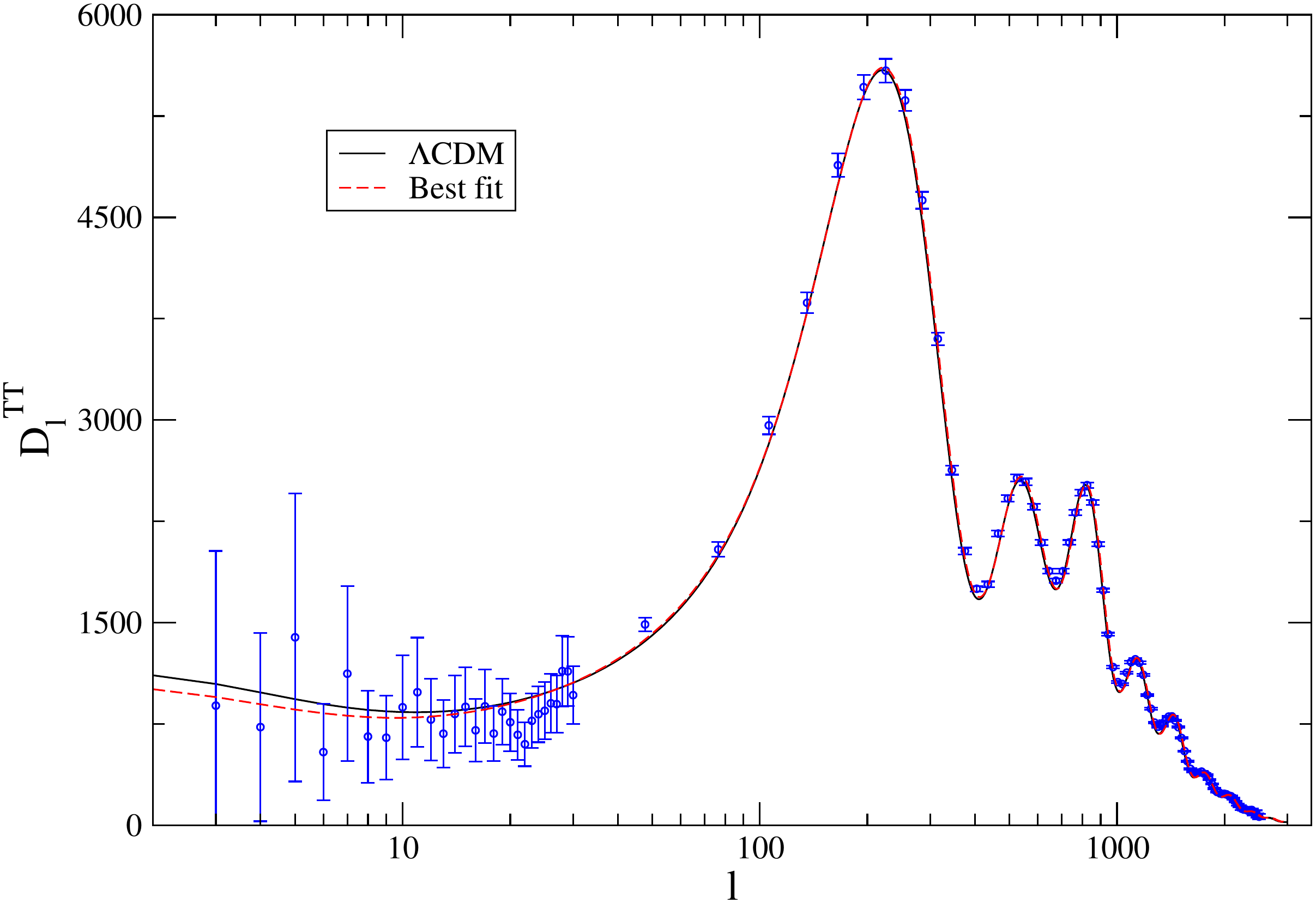} 
	\caption{\label{cmbTT}  {\it{The CMB TT power spectrum 
$D^{TT}_l =l(l+ 1)C_l/2 \pi\mu K^2$, for the interacting scenario at hand (dashed-red 
line) and for the flat $\Lambda$CDM cosmology (black-solid line). The data with their 
error bars have been taken from Planck Collaboration \cite{data_cmb}.}}}
\end{figure}

\begin{figure}
\includegraphics[width=3.3in, height=2.7in]{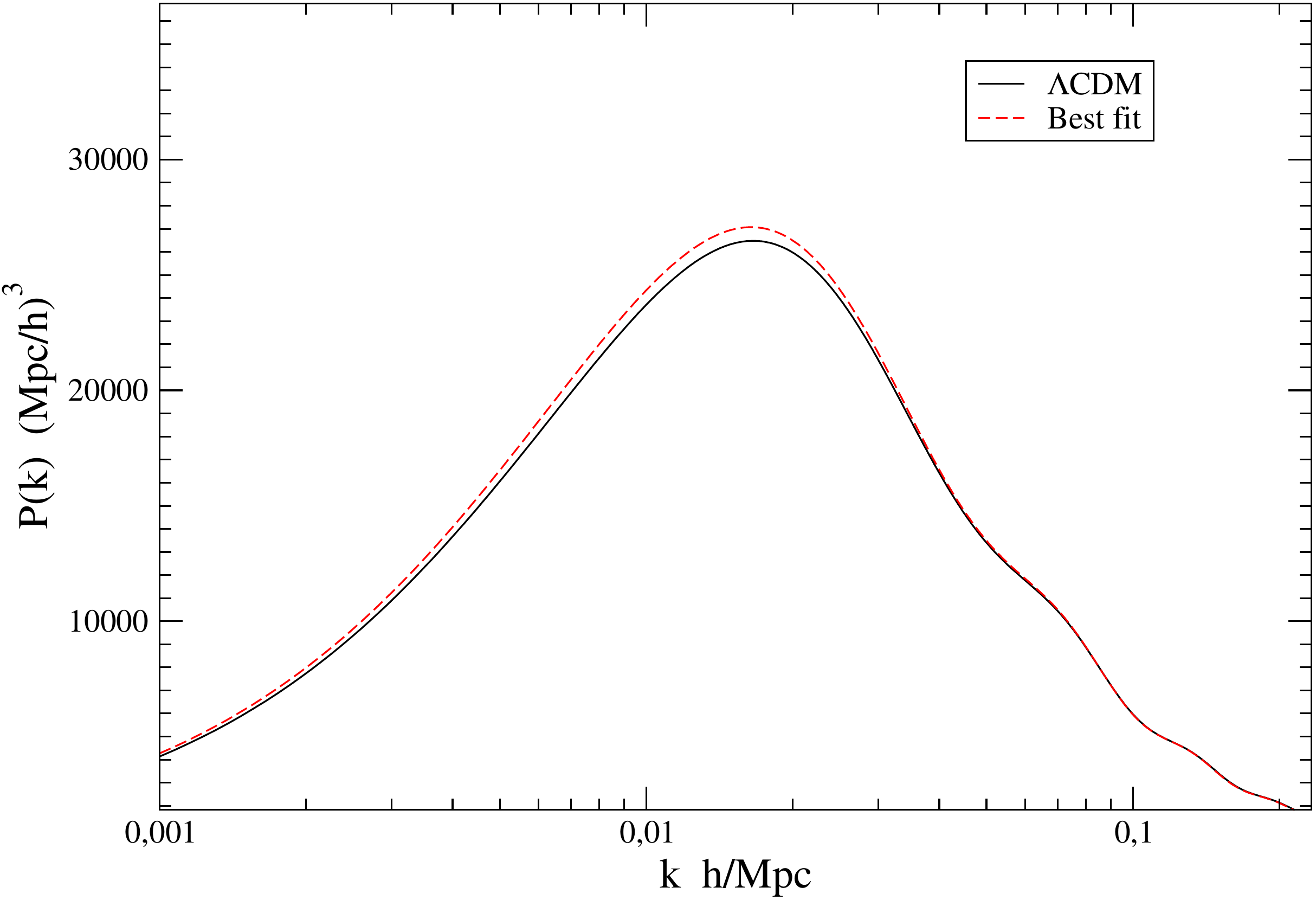} 
\caption{\label{cmb2}\it{The theoretical prediction for linear matter power 
spectrum, for the interacting scenario at hand (dashed-red 
line) and for the flat $\Lambda$CDM cosmology (black-solid line).}}
\end{figure}

 In Figures \ref{cmbTT} and \ref{cmb2} we respectively depict the theoretical 
predictions for the CMB temperature power spectrum and for the linear matter power 
spectrum, for the present interacting dark energy-dark matter model, as well as for the 
flat $\Lambda$CDM scenario. From the CMB temperature power spectrum we deduce that the 
coupling between dark matter - dark energy affects the microwave background temperature 
anisotropies at large angular scales, especially at very low $l$ (in particular for $l < 
30$). Furthermore, significant effects on the linear matter power spectrum are also 
observed at large scales (for $k < 0.034$ h/Mpc). 

\begin{figure*}
\includegraphics[width=5.6in, height=5.6in]{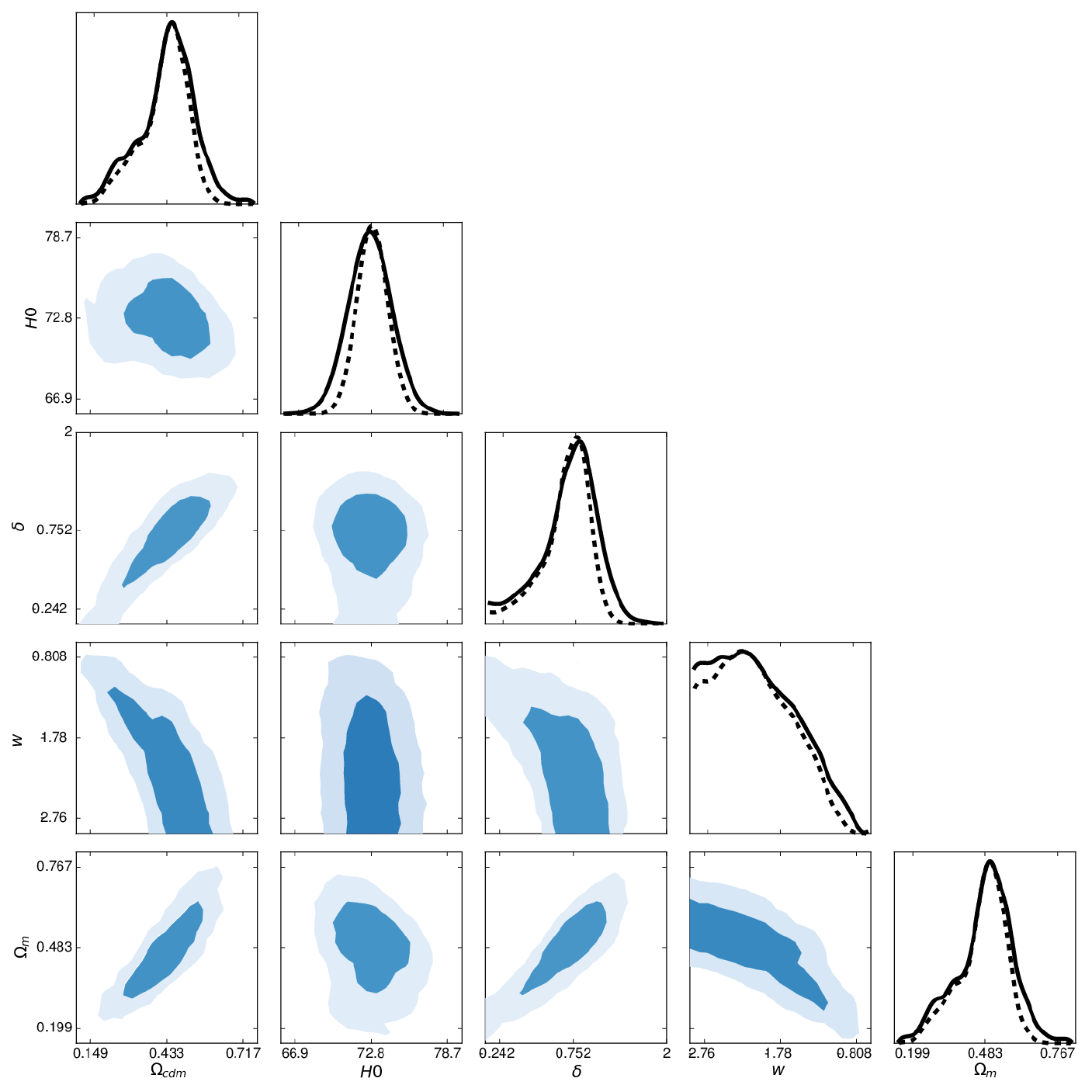}
\caption{\label{Qm1} {\it{ 68.3$\%$ and 95.4$\%$ confidence-level contour plots for the
free parameters of the scenario of interacting dark energy, using only $CC$ + $H_0$
observational data. Additionally, we depict the marginalized one-dimensional posterior
distribution, where the dashed curve stands for the average likelihood distribution.}}}
\end{figure*}

\begin{figure*}
\includegraphics[width=7.0in, height=6.0in]{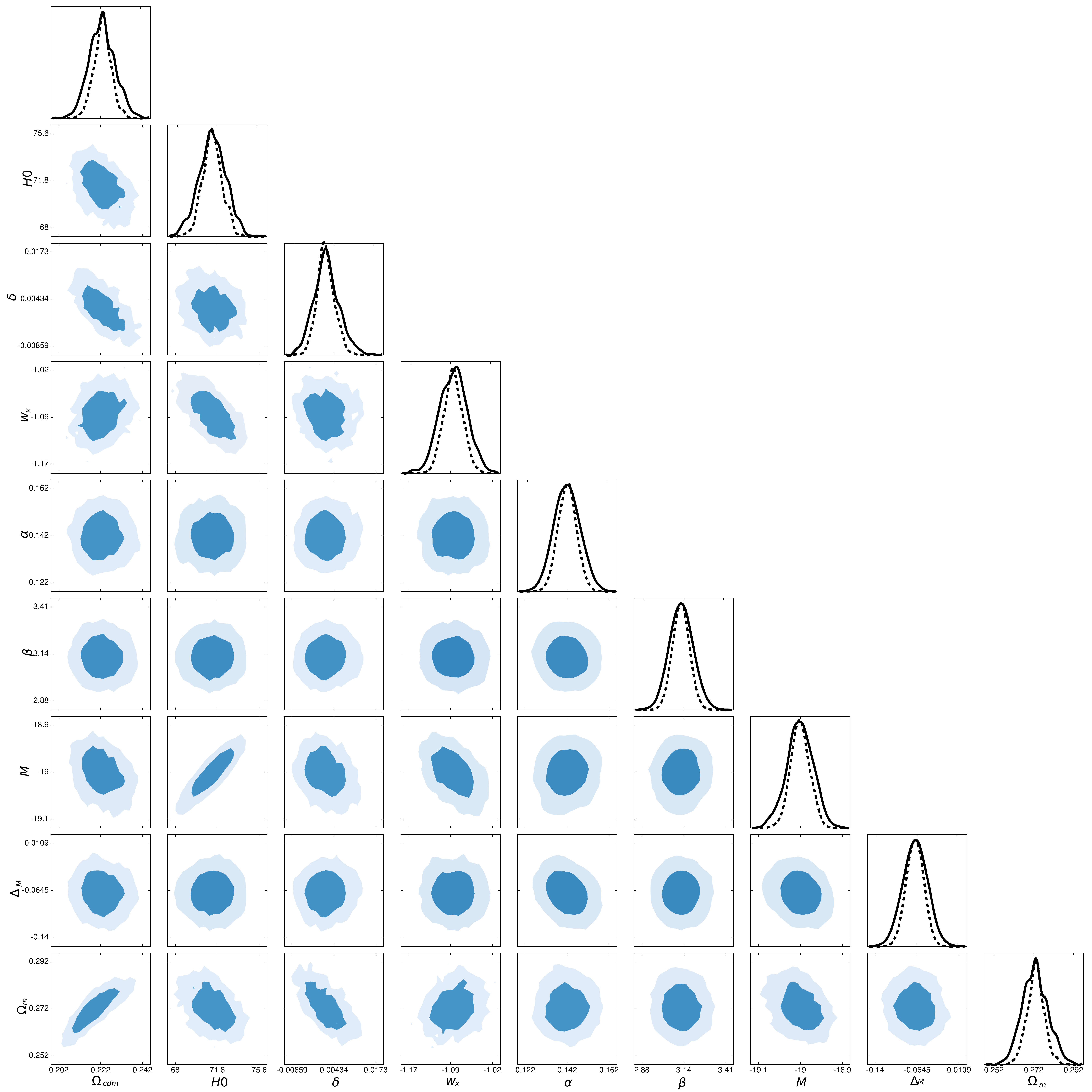}
\caption{\label{Qm2}  {\it{ 68.3$\%$ and 95.4$\%$  confidence-level contour plots for the 
free 
parameters of the
scenario of interacting dark energy, using $CC$ + $H_0$ + SNeIa/JLA + BAO + CMB 
observational
data. Additionally, we depict the marginalized one-dimensional posterior distribution,
where the dashed curve stands for the average likelihood distribution.}}}
\end{figure*}

Additionally, in Tables \ref{tab2} and \ref{tab3} we summarize the main results of our 
statistical analysis, and in Figures \ref{Qm1} and \ref{Qm2} we present the contour plots 
for the free parameters of the scenario of interacting dark energy, using $CC$ + $H_0$ 
and 
$CC$ + $H_0$ +SNeIa/JLA + BAO + CMB
observations, respectively.
\begin{table}[ht]
      \begin{center}
          \begin{tabular}{ccccc}
          \hline
          \hline
           Param. & best-fit & mean$\pm\sigma$ & 95\% lower & 95\% upper \\ \hline
          \hline
$\Omega_{cdm0}$ &$0.386$ & $0.3839_{-0.094}^{+0.098}$ & $0.2045$ & $0.5634$ \\
$H_0$ &$72.67$ & $72.95_{-1.8}^{+2.1}$ & $68.93$ & $76.71$ \\
$\delta$ &$0.4848$ & $0.4749_{-0.24}^{+0.39}$ & $-0.1248$ & $0.9998$ \\
$w$ &$-1.866$ & $-2.004_{-0.58}^{+0.59}$ & $-2.995$ & $-1.142$ \\
$\Omega_{m0}$ &$0.436$ & $0.4339_{-0.094}^{+0.098}$ & $0.2545$ & $0.6134$ \\
          \hline
          \hline
          \end{tabular}
      \end{center}
\caption{Summary of the best fit values and main results for the free parameters of the
scenario of interacting dark energy, using only CC + $H_0$ observational data. The
parameter
$\Omega_{m}$ is the contribution of baryons plus dark matter, namely $\Omega_{m} =
\Omega_{b} + \Omega_{cdm}$.}
      \label{tab2}
\end{table}

 \begin{table}[ht]
      \begin{center}
          \begin{tabular}{ccccc}
          \hline
          \hline
           Param. & best-fit & mean$\pm\sigma$ & 95\% lower & 95\% upper \\ \hline
          \hline
$\Omega_{cdm 0}$ &$0.2246$ & $0.2229_{-0.0069}^{+0.0063}$ & $0.2099$ & $0.2365$ \\ 
$H_0$ &$71.17$ & $71.37_{-1.3}^{+1.3}$ & $68.67$ & $74.01$ \\ 
$\delta$ &$0.00099$ & $0.00196_{-0.0046}^{+0.0038}$ & $-0.00631$ & $0.01085$ \\ 
$w$ &$-1.085$ & $-1.087_{-0.028}^{+0.027}$ & $-1.139$ & $-1.032$ \\ 
$\alpha$ &$0.143$ & $0.1422_{-0.007}^{+0.0065}$ & $0.1291$ & $0.1556$ \\ 
$\beta$ &$3.117$ & $3.126_{-0.083}^{+0.079}$ & $2.966$ & $3.29$ \\ 
$M$ &$-19.04$ & $-19.04_{-0.037}^{+0.041}$ & $-19.12$ & $-18.96$ \\ 
$\Delta_{M}$ &$-0.0721$ & $-0.0680_{-0.023}^{+0.024}$ & $-0.116$ & $-0.0211$ \\ 
$\Omega_{m0}$ &$0.2746$ & $0.2729_{-0.0069}^{+0.0063}$ & $0.2599$ & $0.2865$ \\ 
          \hline
          \hline
          \end{tabular}
      \end{center}
\caption{Summary of the best fit values and main results  for the free parameters of the
scenario of interacting dark energy, using CC $+$ $H_0$  $+$ SNeIa/JLA
$+$ BAO $+$ CMB observational data. The parameters $\alpha$, $\beta$, $M$,
and $\Delta_{M}$ are nuisance parameters as explained in subsection \ref{snia-data}.}
\label{tab3}
\end{table}

As we can see, when analyzing the scenario with $CC$ + $H_0$ data, we observe that
$\delta \neq 0$ combined with  $w < - 1$ is preferred (see Table \ref{tab2}). 
On the other hand, when we break the degeneracy of the parameters,  combining the 
analysis 
with SNeIa/JLA + BAO + CMB 
observations, namely using $CC$ + $H_0$ + SNe Ia $+$ BAO + CMB data, we observe that 
$\delta \simeq 
0$ (with a minor
tendency for $\delta \gtrsim 0$, i.e. for an energy flow from dark energy to dark 
matter), 
however 
with $w < -1$ up to 3$\sigma$ 
confidence level. More specifically we note that $-1.169 \leq w \leq -1.009$ at 
3$\sigma$. 
The 2$\sigma$ confidence-level  bounds can be seen in 
Table \ref{tab3}. 

 Hence, we deduce that the global fits from the map of the universe  
 expansion history have a slight preference towards a cosmological scenario of 
interacting 
 dark energy. This is the main 
 result of the present investigation, and it is in agreement with
 the results of other observational works
 \cite{Salvatelli:2014zta,Sola:2016_b,Sola:2016jky,Sola:2015,He:2010im,Xu:2013jma,
 Costa:2013sva,Ruiz:2014hma,Cai:2015zoa,Li:2015vla,
 Richarte:2015maa,Valiviita:2015dfa, Eingorn:2015rma,Pan:2012ki,Murgia:2016ccp}, however 
it  has been arisen through the novel use of the recently released cosmic chronometers 
data.

\section{Final remarks}
\label{discu}

In the present work we have extracted observational constraints on the scenario of
interacting dark energy, without the need to consider a specific interaction form, using
the recently released cosmic chronometers data. In particular,
considering a general direct interaction between dark matter and dark energy sectors has
the general effect of altering the evolution of the former to $\rho_{cdm} \propto a^{-3+
\delta}$, where $\delta$ is the parameter that quantifies the deviation from the
non-interacting case.
Hence, one can use observational data in order to fit  $\delta$  as well as the other
cosmological parameters.

In our analysis we used (i) the very recently released cosmic chronometer data sets along
with the very latest measured value of the local Hubble parameter, $H_0 = 73.02 \pm 1.79$
km/s/Mpc \cite{riess}, (ii) ``joint light curves'' (JLA) sample containing
740 latest Supernovae Type Ia data points, (iii) baryon acoustic oscillation data points, 
 (iv) CMB distance priors adopting 2015 Planck TT, TE, EE + lowP data, 
marginalized 
over the 
amplitude of the lensing power 
spectrum \cite{cmb4}. We presented two different sets of values of the model parameters, 
one  
arising using the combined set of $CC$ + $H_0$ data (summarized in  Fig. \ref{Qm1} and 
Table
\ref{tab2}), and one arising using the $CC$ + $H_0$+ SNeIa/JLA + BAO + CMB data 
(summarized in Fig. \ref{Qm2} and Table \ref{tab3}).

 We found that the combined analysis $CC$ + $H_0$ slightly favors a non-zero value 
for the interacting parameter $\delta$, while the dark energy equation-of-state parameter 
lies below  $-1$. Additionally, for the combined analysis, using $CC$ + $H_0$+SNeIa/JLA 
$+$ 
BAO $+$ CMB data, we found that $\delta$ is close to zero, nevertheless the possibility 
of a small interaction is not ruled out, and moreover we find that the dark energy 
equation-of-state parameter lies   below  $-1$ at 3$\sigma$ confidence level.

In summary, using the latest observational data, we have found that an interaction 
between dark matter and dark energy sectors is mildly favored.

\section*{Acknowledgments}
S.P. acknowledges Science and Engineering Research Board (SERB), Govt. of India, for 
awarding the National Post-Doctoral Fellowship (File No: PDF/2015/000640). This article 
is 
based upon work from COST Action ``Cosmology and Astrophysics Network
for Theoretical Advances and Training Actions'', supported by COST (European Cooperation
in Science and Technology). The authors thank N. Tamanini for useful comments. Finally, 
the authors 
gratefully acknowledge an anonymous referee for his/her valuable comments.

\end{document}